\documentclass[twocolumn,preprintnumbers,amsmath,amssymb,superscriptaddress]{revtex4}

\usepackage{graphicx}
\begin{document}
\bibliographystyle{prsty}

\title{Digging up bulk band dispersion buried under a passivation layer
}

\author{Masaki~Kobayashi}
\affiliation{Department of Applied Chemistry, 
School of Engineering, University of Tokyo, 
7-3-1 Hongo, Bunkyo-ku, Tokyo 113-8656, Japan}
\affiliation{Swiss Light Source, Paul Scherrer Institut, 
CH-5232 Villigen PSI, Switzerland}
\author{Iriya~Muneta}
\affiliation{Department of Electrical Engineering and Information Systems, 
University of Tokyo, 7-3-1 Hongo, Bunkyo-ku, Tokyo 113-8656, Japan}
\author{Thorsten~Schmitt}
\affiliation{Swiss Light Source, Paul Scherrer Institut, 
CH-5232 Villigen PSI, Switzerland}
\author{Luc~Patthey}
\affiliation{Swiss Light Source, Paul Scherrer Institut, 
CH-5232 Villigen PSI, Switzerland}
\author{Sinobu~Ohya}
\affiliation{Department of Electrical Engineering and Information Systems, 
University of Tokyo, 7-3-1 Hongo, Bunkyo-ku, Tokyo 113-8656, Japan}
\author{Masaaki~Tanaka}
\affiliation{Department of Electrical Engineering and Information Systems, 
University of Tokyo, 7-3-1 Hongo, Bunkyo-ku, Tokyo 113-8656, Japan}
\author{Masaharu~Oshima}
\affiliation{Department of Applied Chemistry, 
School of Engineering, University of Tokyo, 
7-3-1 Hongo, Bunkyo-ku, Tokyo 113-8656, Japan}
\author{Vladimir~N.~Strocov}
\affiliation{Swiss Light Source, Paul Scherrer Institut, 
CH-5232 Villigen PSI, Switzerland}
\date{\today}

\begin{abstract}
Atomically controlled crystal growth of thin films has established foundations of nanotechnology aimed at the development of advanced functional devices. Crystallization under non-equilibrium conditions allows engineering of new materials with their atomically-flat interfaces in the heterostructures exhibiting novel physical properties. From a fundamental point of view, knowledge of the electronic structures of thin films and their interfaces is indispensable to understand the origins of their functionality which further evolves into realistic device application. 
In view of extreme surface sensitivity of the conventional vacuum-ultraviolet (VUV) angle-resolved photoemission spectroscopy (ARPES), with a probing depth of several angstroms, experiments on thin films have to use sophisticated $in$-$situ$ sample transfer systems to avoid surface contamination. 
In this Letter, we put forward a method to circumvent these difficulties using soft X-ray (SX) ARPES. 
A GaAs:Be thin film in our samples was protected by an amorphous As layer with an thickness of $\sim 1$ nm exceeding the probing depth of the VUV photoemission with photon energy $h\nu$ around 100 eV. 
The increase of the probing depth with increasing $h\nu$ towards the SX region has clearly exposed the bulk band dispersion without any surface treatment even after exposure to air. 
Any contributions from potential interface states between the thin film and the amorphous capping layer has been below the detection limit. 
Our results demonstrate that SX-ARPES enables the observation of coherent three-dimensional band dispersion of buried heterostructure layers through an amorphous capping layer, breaking through the necessity of surface cleaning of thin film samples. 
Thereby, this opens new frontiers in diagnostics of authentic momentum-resolved electronic structure of protected thin-film heterostructures. 
\end{abstract}


\maketitle

Development of modern electronics technology requires minimization of the size of electronic devices to the nano or sub-nano scale. Thin-film growth techniques, such as molecular beam epitaxy and pulsed laser deposition, combined with reflection high-energy electron diffraction (RHEED) monitoring, enable fabrication of well-defined crystalline layers with sub-monolayer accuracy. Non-equilibrium growth conditions allow realization of impurity doping levels above the thermal-equilibrial solubility limit. 
The thin-film crystal growth techniques, applied to fabrication of a wide range of electric devices, set up a new avenue for future electronic technology, for instance, oxide-based electronics \cite{NatMater_05_Tsukazaki, NatMater_10_Tsukazaki, Science_10_Mannhart} and spintronics (spin + electronics) \cite{Science_01_Wolf, SST_05_Fukumura}.

In addition to the device application, thin-film crystals are also important from a fundamental point of view because quantum mechanical effects become pronounced on the nano scale, giving rise to novel physical properties which can be well-controlled artificially. Particularly interesting are thin film systems incorporating $3d$ transition-metal elements. 
For example, the interface between band insulator SrTiO$_3$ and LaAlO$_3$ exhibits in-plane metallic conductivity \cite{Nature_04_Ohtomo}, ferromagnetism \cite{PRL_11_Dikin}, and superconductivity at low temperature \cite{Science_10_Reyren}. 
Diluted magnetic semiconductors (DMS) are formed by a semiconductor host doped with magnetic impurities over the solubility limit under non-equilibrium growth conditions. Some of DMS, e.g., Mn-doped GaAs \cite{APL_96_Ohno} and Co-doped TiO$_2$ \cite{Science_01_Matsumoto}, show ferromagnetic properties in spite of only several percent of the magnetic impurity concentration. 
The unprecedented physical properties of the thin films or at their interfaces of heterostructures can be artificially tuned by carrier doping \cite{Nature_00_Ohno}, strain imposed by the substrate \cite{Nature_04_Haeni}, and confinement in heterostructures/superlattices \cite{Science_11_Yoshimatsu, APL_99_Ohtomo}.

Knowledge of the electronic structure is a key element in understanding their rich and colourful physics. Angle-resolved photoemission spectroscopy (ARPES) is one of the most powerful experimental methods to investigate the electronic structure of crystalline solids based on the fact that the photoelectron current dependence on energy and the emission angle reflects the band dispersion as functions of crystal momentum $\mathbf{k}$ in the reciprocal space. 
The conventional ARPES in the vacuum-ultraviolet region (VUV-ARPES) is characterized by a probing depth of several {\AA}, which makes it overly sensitive to the surface immediately affected by the oxidization and contaminations upon exposure to air. Therefore, VUV-ARPES experiments on thin films usually employ an $in$-$situ$ sample transfer from the growth chamber into the analysis one \cite{RSI_03_Horiba}. 
This technique is also effective for three-dimensional materials such as perovskite oxides having no natural cleavable planes. However, this method suffers from technical difficulties of vacuum coupling between the growth and analysis equipment and of characterization of the macroscopic physical properties of the grown thin film before the measurement \cite{PRB_06_Chikamatsu}. 
On the other hand, $ex$-$situ$ VUV-ARPES measurements on thin films will require special surface cleaning, e.g., ion sputtering and annealing, after exposure to air to restore the authentic surface, which is often impossible without its destruction. 
Here, we use SX-ARPES to perform $\mathbf{k}$-resolved electronic structure measurements on thin films protected from oxidation by an amorphous passivation layer, and propose practical applications of this method for various thin film systems.

Although the energy resolution of SX-ARPES, typically up to a few tens of meV, does not match that of the state-of-the-art VUV-ARPES, it has important advantages allowing one to explore various physical properties complementary to the characterization by VUV-ARPES: 
{\bf I}. Long probing depth expressed by the photoelectron inelastic mean free path $\lambda$. According to the `universal' curve \cite{SIA_79_Seah}, increase of $h\nu$ towards 1 keV pushes $\lambda$ above $\sim 1$ nm; 
{\bf II}. Higher intrinsic resolution in surface-perpendicular momentum $k_{\perp}$ determined by the photoelectron spatial confinement within $\lambda$ \cite{JESRP_03_Strocov}; 
{\bf III}. Free electron-like final state achieved due to the high kinetic energy of photoelectrons; 
{\bf IV}. Simplified matrix elements. The advantages {\bf II}-{\bf IV} are significant to observe bulk three-dimensional band dispersion; 
{\bf V}. Resonant phenomena at absorption edges. Monitoring of resonant enhancement of the valence bands enables determination of the element-specific band structure \cite{PRL_08_Im, PRB_10_Mulazzi}. 
In this Letter, we will show how the advantage of SX-ARPES with larger $\lambda$ authorizes access to protected thin films.

\begin{figure}[b!]
\begin{center}
\includegraphics[width=8.8cm]{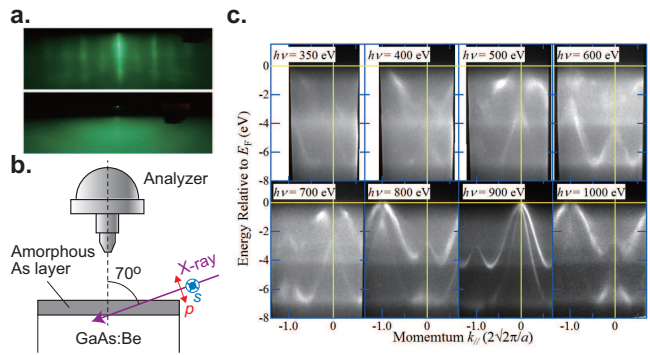}
\caption{ARPES images measured through the amorphous As capping layer on the GaAs thin film. 
(a) RHEED pattern of the GaAs thin film before (top) and after (bottom) As deposition. The characteristic diffraction pattern from the GaAs layer disappears after the As deposition. 
(b) Schematic image of the experimental setup. The analyzer slit lies in the scattering plane including the incident-beam wave vector. Directions of the $p$- and $s$-polarizations are also shown. 
(c) ARPES images taken at various incident photon energy. Here, the Fermi level ($E_\mathrm{F}$) almost corresponds to the valence-band maximum (VBM). These data were taken with $p$-polarization and the sample orientation is along the $\Gamma$-K-X line. 
}
\label{GaAs_AmorAs}
\end{center}
\end{figure}

Our sample of a GaAs:Be thin film with a thickness of 50 nm was homoepitaxially grown on a GaAs(001) substrate at 600 $^{\circ}$C by molecular beam epitaxy method. To avoid surface oxidation, the sample was covered by an amorphous As capping layer with a thickness of $\sim 1$ nm to produce a structure As/GaAs:Be/GaAs(buffer)/GaAs(001). 
It has been reported that such passivation layer practically eliminates carrier depletion on the surface, and protects the surface from the air, water, and heat treatment up to 180 $^{\circ}$C \cite{EL_84_Kawai}. 
The amorphous As layer completely quenched the sharp reflection of the RHEED (reflection of high energy electron diffraction) pattern characteristic of the loss of the system crystallinity, as shown in Fig.~\ref{GaAs_AmorAs}(a). 
The hole concentration of the sample was $\sim 1.0 \times 10^{18}$ cm$^{-3}$ as determined by the Hall effect. 
The soft X-ray ARPES experiments were conducted at the SX-ARPES end station of the ADRESS beamline at the Swiss Light Source \cite{JSR_10_Strocov}. 
The setup of the SX-ARPES measurements is schematically shown in Fig.~\ref{GaAs_AmorAs}(b). 
Here, the incident angle of the beam is 70 degrees to the surface normal. 
The measurements used linear-vertically and linear-horizontally polarized X-rays corresponding to the $p$- and $s$-polarizations, respectively, and were performed under ultrahigh vacuum of $5.0 \times 10^{-11}$ mbar. 
The photoelectron analyzer is PHOIBOS-150 (SPECS Co. Ltd). 
The analyzer slit was oriented in the scattering plane. 
In the $h\nu$ range from 350 to 1000 eV, the combined beamline and analyzer energy resolution including the thermal broadening varied from 50 to 150 meV.

Figure~\ref{GaAs_AmorAs}(c) shows the $h\nu$ dependence of the ARPES images. 
The ARPES image taken $h\nu = 350$ eV is nearly non-dispersive, indicating the loss of the coherent GaAs signal. With increasing $h\nu$, astonishingly, the dispersive features start to pile up distinctly due to coherent photoelectron escaping through the As layer. 
At $h\nu = 900$ eV the coherent intensity is already sufficient for ARPES measurements with an accumulation time of only few minutes per image. 
The evolution of image with $h\nu$ is consistent with the increase of $\lambda$ according to the `universal' curve \cite{SIA_79_Seah}. 
The increase of $\lambda$ can be also confirmed in the Ga $3d$ core-level spectra (not shown).
Further variation of the ARPES images as a function of $h\nu$ comes from the $k_z$ dependence of the band structure.

\begin{figure*}[t!]
\begin{center}
\includegraphics[width=16cm]{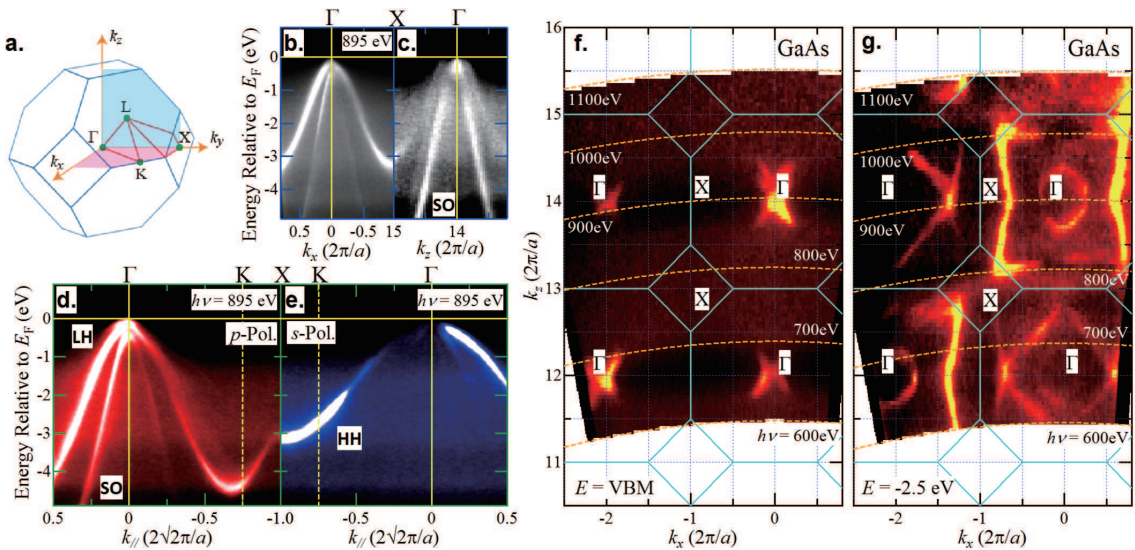}
\caption{Valence-band dispersion of the GaAs thin film. 
(a) Brillouin zone (BZ). The pink and blue areas are for the surface-perpendicular ($k_x$-$k_y$) and the surface-normal ($k_z$-$k_{\parallel}$) plots, respectively. 
(b),(c) Energy($E$)-momentum plots of the $\Gamma$-X-$\Gamma$ symmetric line along the surface-perpendicular $k_x$ and surface-normal $k_z$ momenta, respectively. SO denotes the split-off band of GaAs. 
(d),(e) $E$-$k_{\parallel}$ plots of the $\Gamma$-K-X symmetric line taken with $p$- and $s$-polarizations, respectively. The dashed vertical lines are the BZ boundary at the K point. LH and HH denote the light-hole and heavy-hole bands, respectively. 
(f),(g) Surface-normal constant-energy mappings in vicinity of VBM and $E = -2.5$ eV, respectively. Energy window (integrated range) for the mappings is $\pm 50$ meV. 
The dashed lines represent the photoelectron momentum in reciprocal space estimated from $h\nu$ and emission angle. 
The solid lines denote boundaries of the BZ. 
}
\label{GA_Ek}
\end{center}
\end{figure*}

The ARPES signal from the amorphous As layer is non-dispersive and contributes only to the (energy dependent) spectral background in the dispersive ARPES image. 
The appearance of the dispersive band structure from the underlayer is non-trivial because, seemingly, the As layer does not scatter photoelectrons. 
The effect is explained, first of all, by that elastic scattering by the As atoms at random positions results in homogeneous photoelectron scattering to form only a $\mathbf{k}$-averaged background. 
There are no signs of superstructures or additional $\mathbf{k}$-broadening of the dispersive features characteristic of inter-atomic correlations in the protective layer. 
Inelastic scattering in the amorphous As layer is expected to be weak at high photon energy and also contributes only to the $\mathbf{k}$-independent background.

The electronic structure, i.e., the band dispersion $E(k)$ of the GaAs underlayer is evident from the energy-momentum plots of the ARPES intensity. 
The whole $E(k)$ of GaAs along the $\Gamma$-X-$\Gamma$ symmetric line in the surface-parallel momentum direction $k_x$ is identical to that in the surface-perpendicular momentum direction $k_z$ obtained under variation of $h\nu$, as shown in Figs.~\ref{GA_Ek}(b) and \ref{GA_Ek}(c), reflecting the three-dimensionality of the band structure. 
The larger broadening in the $k_z$-plot compared to the $k_{\parallel}$-plot in Fig.~\ref{GA_Ek}(b) manifests the $k_z$-broadening which is due to the photoelectron confinement within the mean free path $\lambda$ \cite{JESRP_03_Strocov}. 
The difference of the intensity in the band dispersion comes from matrix element effects. 
Figures~\ref{GA_Ek}(d) and \ref{GA_Ek}(e) show the linear-polarization dependence of the ARPES image of $E(k)$. 
The light-hole and split-off band dispersion along the $\Gamma$-K-X line can be seen in the ARPES image taken with $p$-polarization, as shown in Fig.~\ref{GA_Ek}(d). In contrast, the heavy-hole band appears in the ARPES image with $s$-polarization, as shown in Fig.~\ref{GA_Ek}(e). 
The energy-momentum plots reveal the characteristic band structure of GaAs \cite{PRB_88_Cardona, PRL_09_Luo}. 
Figures~\ref{GA_Ek}(f) ~\ref{GA_Ek}(g) show the constant-energy mapping (CEM) in vicinity of the valence-band maximum (VBM) and $E = -2.5$ eV, respectively, where VBM almost corresponds to the Fermi level ($E_\mathrm{F}$). 
The ARPES data along the $k_z$ direction were measured by varying of $h\nu$. 
Its periodicity is consistent with the Brillouin zone (BZ) sequence. 
The intensity in vicinity of VBM around the $\Gamma$ point of the 7th BZ ($h\nu = 895$ eV) is slightly higher than that of the 6th BZ ($h\nu = 655$ eV), perhaps owing to the increase of $\lambda$ with $h\nu$, slight depletion of the hole-carrier concentration near the interface and/or modulation of the photoemission intensity due to the matrix element effects. 
Here, we make the point that SX-ARPES on a heterostructure thin film has enabled a transparent observation of bulk three-dimensional electronic structure of the underlayer through the amorphous capping layer without any surface treatment.

The experimental data show no signs of interface states which may exist between the GaAs and the capping layers. Indeed, such states would be two-dimensional, with the absence of dispersion in the $k_{\perp}$ direction which would be manifested as vertical lines in the $k_{\perp}$ dependent CEM of Fig.~\ref{GA_Ek}(f). 
As no such structure is seen, we conclude that either interface states are not induced by the amorphous layer, or contribution from the interface/surface states to ARPES images is vanishingly small at high energy \cite{JAP_09_Kobayashi, PRL_12_Elia}.

\begin{figure}[t!]
\begin{center}
\includegraphics[width=8.8cm]{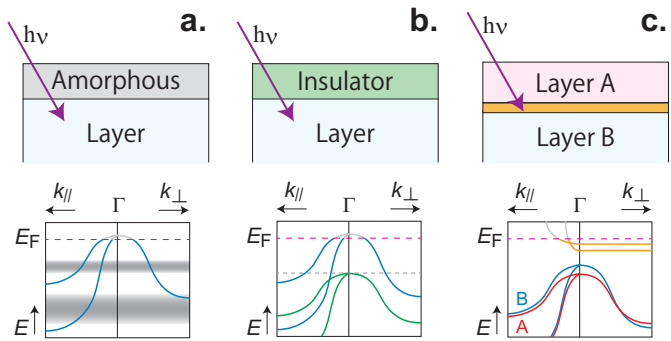}
\caption{Application of SX-ARPES to various thin-film systems. The bottom panels show schematic band dispersion around $\Gamma$ point along the surface-parallel $k_{\parallel}$ and surface-perpendicular momentum $k_{\perp}$ directions. 
(a) Thin film capped by an amorphous overlayer. 
(b) Heterostructure of a metallic underlayer with an insulating overlayer. 
(c) Heterostructure of different insulating compounds with a metallic interface state. 
}
\label{AppSxarpes}
\end{center}
\end{figure}

Based on the experimental findings, we now discuss the perspectives of SX-ARPES applications to various heterostructures. 
Figure~\ref{AppSxarpes} shows schematic diagrams of three such systems: 
{\bf 1}. Thin film covered by an amorphous layer [Fig.~\ref{AppSxarpes}(a)]; 
{\bf 2}. Heterostructure of metallic underlayer and insulating overlayer [Fig.~\ref{AppSxarpes}(b)]; 
{\bf 3}. Interface state between two compounds [Fig.~\ref{AppSxarpes}(c)]. 
In the present Letter, we have demonstrated case 1 of the amorphous passivating overlayer, which produces merely a homogeneous background in the ARPES spectra due to its disordered nature. 
In addition to the protection of the surface from oxidation, adjustment of the amorphous layer composition can be used to induce carrier depletion or doping into the underlayer in proximity of the interface. 
In case 2 shown in Fig.~\ref{AppSxarpes}(b), it will be possible to measure band dispersion of the buried metallic layer through a crystalline insulating overlayer. 
The experimental band dispersion in vicinity of $E_\mathrm{F}$ will come only from the metallic underlayer, because the insulating overlayer has here no density of states. Outside the band gap of the insulating overlayer, both the bands of metallic and insulating compounds coexist, but they can be distinguished by changing $h\nu$ to vary $\lambda$ or using resonance enhancement at the absorption edges. 
Furthermore, SX-ARPES can be useful in case 3, as shown in Fig.~\ref{AppSxarpes}(c), to explore the interface state of the heterostructure such as LaAlO$_3$/SrTiO$_3$ \cite{Nature_04_Ohtomo}. 
Indeed, metallic interface states formed between the band insulator SrTiO$_3$ and the Mott insulator LaTiO$_3$ were successfully observed by resonance soft x-ray (angle-integrated) photoemission \cite{PRL_06_Takizawa}. 
Even if the ARPES signal from the overlayer, interface, and underlayer are entangled, the flat dispersion along the $k_{\perp}$ direction seen in the $h\nu$ dependence will separate the two-dimensional interface contribution from the bulk ones. 
If the underlayer and/or overlayer include $d$ or $f$ elements, resonance ARPES can also be helpful to identify the interface state. 
Additionally, SX-ARPES can monitor the band dispersion in heterostructures tuned by lattice mismatch between the layers, which will shed light on the strain effects on the electronic structure. 
This method will be useful not only for correlated electron heterostructures but also for conventional semiconductor heterostructures, e.g., FET or MOS structures.

The method of SX-ARPES combined with $ex$-$situ$ treatment of protected thin films may be complementary to the conventional VUV-ARPES with $in$-$situ$ treatment. 
The unnecessity of surface cleaning in this method leads to the convenience and high throughput of the ARPES diagnostics, and the passivation layer allows fast screening of various systems to find the most promising materials for more elaborate $in$-$situ$ ARPES. 
As the energy resolving power of the state-of-the-art VUV-ARPES (typically $< 10$ meV) is higher than that of SX-ARPES (order of $\sim 100$ meV), subsequent $in$-$situ$ VUV-ARPES measurements will provide with a more detailed picture of the electronic structure, in particular in the region near $E_\mathrm{F}$ participating in transport properties. 
The idea of ARPES looking through a passivation layer naturally extends to the hard X-ray (HX) ARPES owing to further increase of $\lambda$ with $h\nu$ \cite{NatMater_11_Gray}, although this is connected with a loss of the coherent band structure signal due to destructive effect of electron-phonon interaction. 
In this way, our method of the SX-ARPES on protected heterostructures will deepen the fundamental understandings of the innovative physics of thin-film devices, and promote further development of their science and practical applications.

The authors thank C. Quitmann, F. van der Veen and J. Mesot for their continuous support of the SX-ARPES project at SLS. 
M.K. acknowledges support from the Japan Society for the Promotion of Science.

\end{document}